\documentclass[fleqn,12pt,twoside]{article}
\usepackage{espcrc1}

\usepackage{graphicx}
\usepackage[figuresright]{rotating}

\newcommand{\beq}{\begin{equation}}
\newcommand{\eeq}{\end{equation}}

\newcommand{\AmS}{{\protect\the\textfont2
  A\kern-.1667em\lower.5ex\hbox{M}\kern-.125emS}}

\title{QCD and Heavy Ions}

\author{D. Kharzeev\address{Physics Department, \\
         Brookhaven National Laboratory, \\
         Upton, New York 11973-5000, USA}
\thanks{Work supported by the US Department of Energy under contract No DE-AC02-
98CH10886.}}

\begin{document}

\maketitle

\begin{abstract}

This short paper is an attempt to describe a theorist's view of the goals  
of relativistic heavy ion program which has just entered the collider era.
These goals are centered around understanding the properties and the 
critical behavior of Quantum Chromo--Dynamics (QCD) 
in the non--linear regime of high color field strength and high 
parton density.    
Some of the current theoretical challenges are highlighted, 
and the place of heavy ion research in the broader 
context of modern particle and nuclear physics is discussed.  

\end{abstract}

\section{WHAT IS QCD?}

Strong interaction is, indeed, the strongest force of Nature. 
It is responsible for over $80 \%$ of the baryon masses, and thus 
for most of the mass of everything on Earth 
and in the Universe. Strong interactions bind  
nucleons in nuclei, which, being then bound into molecules by 
much weaker electro-magnetic forces, 
give rise to the variety of the physical World.
Quantum Chromo--Dynamics is {\it the} theory of strong interactions, 
and its practical importance is thus undeniable. 
But QCD is more than a useful tool -- it is a consistent and 
very rich field theory, which continues to serve as a stimulus for, 
and testing ground of, many exciting ideas and new methods in 
theoretical physics.

\subsection{The structure of QCD}

So what is QCD? From the early days of the accelerator experiments 
it has become clear that the number of hadronic resonances 
is very large, suggesting that all hadrons may be 
classified in terms of a smaller number of (more) 
fundamental constituents. A convenient classification was offered 
by the quark model, but QCD was not born until the hypothetical 
existence of quarks was not supplemented by the principle of 
local gauge invariance, previously established as the basis 
of electromagnetism.  The resulting Lagrangian has the form 
\beq
{\cal{L}} = -{1 \over 4} G_{\mu\nu}^a G_{\mu\nu}^a + \sum_f \bar{q}_f^a 
(i \gamma_{\mu} D_{\mu} - m_f) q_f^a; \label{lagr}
\eeq
the sum is over different colors $a$ and quark flavors $f$; 
the covariant derivative is $D_{\mu} = \partial_{\mu} - i g 
A_{\mu}^a t^a$, where $t^a$ is the generator of the color group $SU(3)$, $A_{\mu}^a$ 
is the gauge (gluon) field and $g$ is the coupling constant. 
The gluon field strength tensor is given by 
\beq
G_{\mu\nu}^a = \partial_{\mu} A_{\nu}^a - \partial_{\nu} A_{\mu}^a + 
g f^{abc} A_{\mu}^b A_{\nu}^c, \label{ftens}
\eeq
where $f^{abc}$ is the structure constant of $SU(3)$: 
$[t^a, t^b] = i f^{abc} t^c$. 

\subsection{Asymptotic freedom}

\underline{\em{Screening and anti--screening of charge.}}
Due to the quantum effects of vacuum polarization, the charge  
in field theory can vary with the distance. In electrodynamics, summation 
of the electron--positron loops in the photon propagator leads 
to the following expression for the effective charge, valid 
at $r \gg r_0$: 
\beq
\alpha_{em}(r) \simeq {3 \pi \over 2 \ln(r/r_0)}. \label{aem}
\eeq
This formula clearly exhibits the ``zero charge'' problem \cite{Landau} 
of QED: in the local 
limit $r_0 \to 0$ the effective charge vanishes at any finite 
distance away from the bare charge due to the screening. 
Fortunately, because of the smallness of the physical coupling, 
this apparent inconsistency of the theory 
manifests itself only at very short distances 
$\sim exp\{-3\pi/[2 \alpha_{em}]\}, \ \alpha_{em} \simeq 1/137$. 
Such short distances are (and probably will 
always remain) beyond the reach of experiments, and one can 
safely use QED as a truly effective theory.

As it has been established long time ago \cite{asfr}, QCD is drastically 
different from electrodynamics in possessing the remarkable property 
of ``asymptotic freedom'' -- due to the 
fact that gluons carry color, the behavior 
of the effective charge $\alpha_s = g^2/4 \pi$ 
changes from the familiar from QED screening to anti--screening:
\beq 
\alpha_{s}(r) \simeq {3 \pi \over (11 N_c / 2 - N_f) \ln(r_0/r)}; 
\label{as} 
\eeq 
as long as the number of flavors does not exceed $16$ ($N_c = 3$), 
the anti--screening 
originating from gluon loops overcomes the screening due to quark--antiquark 
pairs, and the theory, unlike electrodynamics, is weakly coupled 
at short distances: $ \alpha_{s}(r) \to 0$ when $r \to 0$. 

\underline{\em{Does $\alpha_s$ ever get large?}}   
Asymptotic freedom ensures the applicability of QCD perturbation theory 
to the description of processes accompanied by high momentum transfer $Q$. 
However, as $Q$ decreases, the strong coupling $\alpha_s(Q)$ 
grows, and the convergence of perturbative series is lost. 
How large can $\alpha_s$ get? 
The analyzes of many observables suggest that the QCD coupling 
may be ``frozen'' in the infrared region at the value 
$\left<\alpha_s\right>_{IR} \simeq 0.5$ (see 
\cite{Yuri} and references therein). 
Gribov's program \cite{Gribov} relates the freezing of the coupling constant 
to the existence of massless quarks, which leads to the ``decay'' of the vacuum at 
large distances similar to the way it happens in QED in the presence of 
``supercritical'' charge $Z > 1/\alpha$. One may try to  
infer the information about 
the behavior of the coupling constant at large distances by 
performing the matching of the fundamental theory onto the effective chiral 
Lagrangian \cite{FK}. 
The results of \cite{FK} lead to the coupling constant which freezes at the  
value 
\beq
\langle \alpha_s \rangle_{IR} = 
{6 \sqrt{2}\ \pi \over 11 N_c - 2 N_f} \ \sqrt{{N_f^2-1 \over N_c^2-1}};
\eeq 
numerically, for QCD with $N_c=3$ and $N_f=2$ one finds 
$\langle \alpha_s \rangle_{IR} \simeq 0.56$.  
It remains to be seen if a consistent perturbative scheme can be built 
on the basis of this approach \cite{BKM}.

\subsection{Chiral symmetry}

In the limit of massless quarks, QCD Lagrangian (\ref{lagr}) possesses 
an additional symmetry $U_L(N_f) \times U_R(N_f)$ 
with respect to the independent transformation of left-- and right--handed 
quark fields $q_{L,R} = {1 \over 2}(1 \pm \gamma_5) q$: 
\beq
q_L \to V_L q_L; \ \ q_R \to V_R q_R; \ \ V_L, V_R \in U(N_f); \label{chiral}
\eeq
this means that left-- and right--handed quarks are not correlated.  
Even a brief look into the Particle Data tables, or simply in the 
mirror, can convince anyone 
that there is no symmetry between left and right in the physical World. 
One thus has to assume that the symmetry (\ref{chiral}) is spontaneously 
broken in the vacuum.  The flavor composition of the existing eight Goldstone 
bosons (3 pions, 4 kaons, and the $\eta$) suggests that the 
$U_A(1)$ part of $U_L(3) \times U_R(3) = 
SU_L(3) \times SU_R(3) \times U_V(1) \times U_A(1)$ does not exist.  
This constitutes the famous ``$U_A(1)$ problem''. 

\subsection{The origin of mass}

There is yet another problem with the chiral limit in QCD. Indeed, as the 
quark masses are put to zero, the Lagrangian (\ref{lagr}) does not contain 
a single dimensionful scale -- the only parameters are pure numbers $N_c$ 
and $N_f$. The theory is thus apparently 
invariant with respect to scale transformations, 
and the corresponding scale current is conserved: 
$\partial_{\mu} s_{\mu} = 0$.      
However, the absence of a mass scale would imply that all physical 
states in the theory should be massless!

\subsection{Quantum anomalies and classical solutions}

Both apparent problems -- the missing $U_A(1)$ symmetry and the 
origin of hadron masses -- are related to quantum anomalies. 
Once the coupling to gluons is included, both flavor singlet axial 
current and the scale current
cease to be conserved; their divergences 
become proportional to the $\alpha_s G_{\mu\nu}^a \tilde{G}_{\mu\nu}^a$ and 
$\alpha_s G_{\mu\nu}^a G_{\mu\nu}^a$ gluon operators, correspondingly. 
This fact by itself would not have dramatic consequences 
if the gluonic vacuum were ``empty'', with $G_{\mu\nu}^a = 0$. 
However, it appears that due to non--trivial topology of the $SU(3)$ 
gauge group, QCD equations of motion allow classical solutions even 
in the absence of external color source, i.e. in the vacuum. 
The well--known example of a classical solution is the instanton, 
corresponding 
to the mapping of a three--dimensional sphere $S^3$ into the $SU(2)$ subgroup 
of $SU(3)$; its existence was shown to solve the $U_A(1)$ problem.

\subsection{Confinement}

The list of the problems facing us in the study of QCD would not be complete 
without the most important problem of all -- why are the colored quarks and 
gluons excluded from the physical spectrum of the theory? 
Since confinement does not appear in perturbative treatment of the theory, 
the solution of this problem, again, must lie in the properties of the 
QCD vacuum.

\subsection{Understanding the Vacuum}

As was repeatedly stated above, the most important problem facing us in 
the study of all aspects of QCD is understanding the structure of the 
vacuum, which, in a manner of saying, does not at all behave as 
an empty space, but as a physical entity with a complicated structure. 
As such, the vacuum can be excited, altered and modified in physical 
processes \cite{td}; this brings us to the main topic of this talk.

\section{WHY STUDY QCD WITH HEAVY IONS?}

Most of the applications of QCD so far have been  
limited to the short distance regime of high momentum transfer, 
where the theory becomes weakly coupled and can be linearized.
While this is the only domain where our theoretical tools based 
on perturbation theory are adequate, this is also the domain in 
which the beautiful non--linear structure of QCD does not yet reveal 
itself fully. On the other hand, as soon as we decrease the momentum 
transfer in a process, the dynamics rapidly becomes non--linear, but our  
understanding is hindered by the large coupling. 
Being perplexed by this problem, one is 
tempted to dream about an environment in which the coupling is weak, 
allowing a systematic theoretical treatment, but the fields are strong, 
revealing the full non--linear nature of QCD. 
I am going to argue now that this environment can be created on Earth 
with the help of relativistic heavy ion colliders.    

Let us consider an external probe $J$ interacting with the 
nuclear target of atomic number $A$. At small values of Bjorken $x$, 
by uncertainty principle the interaction develops over large 
longitudinal distances $z \sim 1/mx$, where $m$ is the 
nucleon mass. As soon as $z$ becomes larger than the nuclear diameter, 
the probe cannot distinguish between the nucleons located on the front and back edges 
of the nucleus, and all partons within the transverse area $\sim 1/Q^2$ 
determined 
by the momentum transfer $Q$ participate in the interaction coherently. 
The density of partons in the transverse plane is given by
\beq
\rho_A \simeq {x G_A(x,Q^2) \over \pi R_A^2} \sim A^{1/3},
\eeq
where we have assumed that the nuclear gluon distribution  
scales with the number of nucleons $A$. The probe interacts with 
partons with cross section $\sigma \sim \alpha_s / Q^2$; therefore, 
depending on the magnitude of momentum transfer $Q$, atomic number $A$, 
and the value of Bjorken $x$, one may encounter two regimes:
\begin{itemize}
\item{$\sigma \rho_A \ll 1$ -- this is a familiar ``dilute'' regime of 
incoherent interactions, which is well described by the methods of 
perturbative QCD;}
\item{$\sigma \rho_A \gg 1$ -- in this regime, we deal with a dense 
parton system. Not only do the ``leading twist'' expressions become 
inadequate, but also the expansion in higher twists, i.e. in 
multi--parton correlations, breaks down here.}
\end{itemize} 

\begin{figure}[h]
\begin{minipage}[t]{140mm}
\includegraphics[width=14pc]{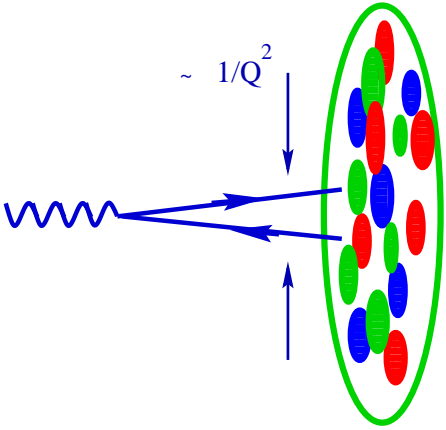}
\caption{Hard probe interacting with the nuclear target 
resolves the transverse distance $\sim 1/\sqrt{Q}$ ($Q^2$ is the square of the 
momentum transfer) and, in the target rest frame, the longitudinal 
distance $\sim 1/(m x)$ ($m$ is the nucleon mass and $x$ -- Bjorken variable).} 

\label{fig:satpict}
\end{minipage}
\end{figure}

The border between the two regimes can be found from the condition 
$\sigma \rho_A \simeq 1$; it determines the critical value of the 
momentum transfer (``saturation scale''\cite{GLR}) at which the parton system 
becomes to look dense to the probe\footnote{Note that since 
$x G_A(x,Q_s^2) \sim A^{1/3}$, 
which is the length of the target, this expression 
in the target rest frame can also be understood as describing a broadening 
of the transverse momentum resulting from the multiple re-scattering 
of the probe.}:
\beq 
Q_s^2 \sim  \alpha_s \ {x G_A(x,Q_s^2) \over \pi R_A^2}. \label{qsat}
\eeq
In this regime, the number of gluons from (\ref{qsat}) is given by 
\beq
x G_A(x,Q_s^2) \sim {\pi \over \alpha_s(Q_s^2)}\ Q_s^2 R_A^2, 
\eeq
where $Q_s^2 R_A^2 \sim A$. 
One can see that the number of gluons 
is proportional to the {\em inverse} of $\alpha_s(Q_s^2)$, and 
becomes large in the weak coupling regime. In this regime,
as we shall now discuss, the dynamics is likely to become 
essentially classical. 

Indeed, the condition (\ref{qsat}) can be derived in the following, 
rather general, way. As a first step, let us re-scale the gluon fields 
in the Lagrangian (\ref{lagr}) as follows: $A_{\mu}^a \to \tilde{A}_{\mu}^a = 
g A_{\mu}^a$. In terms of new fields, $\tilde{G}_{\mu \nu}^a = 
g G_{\mu \nu}^a = \partial_{\mu} \tilde{A}_{\nu}^a - \partial_{\nu} 
\tilde{A}_{\mu}^a +  f^{abc} \tilde{A}_{\mu}^b \tilde{A}_{\nu}^c$, 
and the dependence of the action corresponding to the 
Lagrangian (\ref{lagr}) on the coupling constant is given by  
\beq
S \sim \int {1 \over g^2}\ \tilde{G}_{\mu \nu}^a  \tilde{G}_{\mu \nu}^a 
\ d^4 x. \label{act}
\eeq
Let us now consider a classical configuration of gluon fields; by definition, 
$\tilde{G}_{\mu \nu}^a$ in such a configuration does not depend on 
the coupling, and the action is large, $S \gg \hbar$. The number of 
quanta in such a configuration is then
\beq
N_g \sim {S \over \hbar} \sim {1 \over \alpha_s}\ \rho_4 V_4, \label{numb}
\eeq
where we re-wrote (\ref{act}) as a product of four--dimensional 
action density $\rho_4$ and the four--dimensional volume $V_4$. 
 
The effects of non--linear interactions among the gluons become 
important when $\partial_{\mu} \tilde{A}_{\mu} \sim \tilde{A}_{\mu}^2$ 
(this condition can be made explicitly gauge invariant if we derive it 
from the expansion of a correlation function of gauge-invariant 
gluon operators, e.g., $\tilde{G}^2$). In momentum space, this 
equality corresponds to 
\beq
Q_s^2 \sim \tilde{A}^2 \sim (\tilde{G}^2)^{1/2} = 
\sqrt{\rho_4}; \label{nonlin}
\eeq
$Q_s$ is the typical value of the gluon momentum below which 
the interactions become essentially non--linear. 

Consider now a nucleus $A$ boosted to a high momentum. By uncertainty 
principle, the gluons with transverse momentum $Q_s$ are extended 
in the longitudinal and proper time directions by $\sim 1/Q_s$; 
since the transverse area is $\pi R_A^2$, the four--volume 
is $V_4 \sim \pi R_A^2 / Q_s^2$. The resulting four--density from 
(\ref{numb}) is then 
\beq
\rho_4 \sim \alpha_s\ {N_g \over V_4} \sim \alpha_s\ {N_g\ Q_s^2 
\over \pi R_A^2} 
\sim Q_s^4, \label{class}
\eeq
where at the last stage we have used the non--linearity condition (\ref{nonlin}),  
$\rho_4 \sim Q_s^4$. It is easy to see that (\ref{class}) coincides with the 
saturation condition (\ref{qsat}), since the number of gluons in the 
infinite momentum frame $N_g \sim x G(x,Q_s^2)$. This simple derivation  
illustrates that the physics in the high--density regime can potentially 
be understood in terms of classical gluon fields. 
This correspondence allowed to formulate an effective 
quasi--classical theory \cite{MV}, which is a subject of vigorous  
investigations at present (see, e.g., \cite{KV}).
 
In nuclear collisions, the saturation scale becomes a function of centrality; 
a generic feature of the quasi--classical 
approach -- the proportionality of the number of gluons to the inverse 
of the coupling constant (\ref{numb}) -- thus leads to definite predictions \cite{KN} 
on the centrality dependence of multiplicity, which are so far 
in accord with the data coming from RHIC \cite{data}. The crucial test 
of these ideas will come from the data taken at higher energies, where 
the saturation scale $Q_s^2$ should be larger, and according to the logarithmic 
running of $\alpha_s$ (\ref{as}), the centrality dependence of multiplicity 
should become more flat.     
 
The possible relevance of classical theory raises an interesting question -- 
Weizs\"{a}cker-Williams gluon field of a fast nucleus can be found \cite{MV,KV} from 
the QCD analog of Maxwell equation 
\beq
\partial_{\mu}G_{\mu \nu} = J_{\nu} \label{maxw}
\eeq
with color charges inside the nucleus acting as an external source for gluons. 
On the other hand, QCD possesses classical (Euclidean) 
solutions even in empty space. 
Can new classical solutions to (\ref{maxw}) different 
from Weizs\"{a}cker-Williams fields be found? Does topology play a r\^{o}le in 
relativistic heavy ion collisions? Can topological effects induce  
violations of discrete symmetries manifesting themselves in the 
multi--meson correlations \cite{KPT}? The full answer is lacking 
despite some recent progress \cite{prog,prog1}. 

What happens when relativistic heavy ions (looking like dense ``gluon walls'' 
at sufficiently high energy \cite{Bj}) collide? 
One thing we now know for sure is that collisions at RHIC energies  
produce on the order of a thousand particles per unit of rapidity \cite{data}. 
A system with a number of particles that big can be described by statistical 
methods, and, given the high density of the produced partonic system, an approach 
to equilibrium is likely \cite{therm}, leading to a state of matter known as 
{\em the quark--gluon plasma}. The study of the critical behavior of 
strongly interacting matter and the  
dynamics of phase transitions -- deconfinement, chiral, $U_A(1)$ -- is the 
central goal of the heavy ion program. These topics have been extensively 
discussed in the literature; recent reviews can be found in \cite{revs}. 
The steady progress  
has been driven largely by numerical simulations on the lattice; for a comprehensive 
review, see \cite{latrev}.
The most important problem in the field of heavy ion physics is isolating 
reliable {\em signatures} of collective behavior. Due to the lack of space, 
I cannot dwell on this issue here, and refer the reader to recent reviews \cite{sign} 
which discuss observables in heavy ion collisions. Quarkonium suppression \cite{ms} 
and jet quenching \cite{wg} are among the most promising hard probes of dense 
partonic matter.

\begin{figure}[h]
\begin{minipage}[t]{140mm}
\includegraphics[width=21pc]{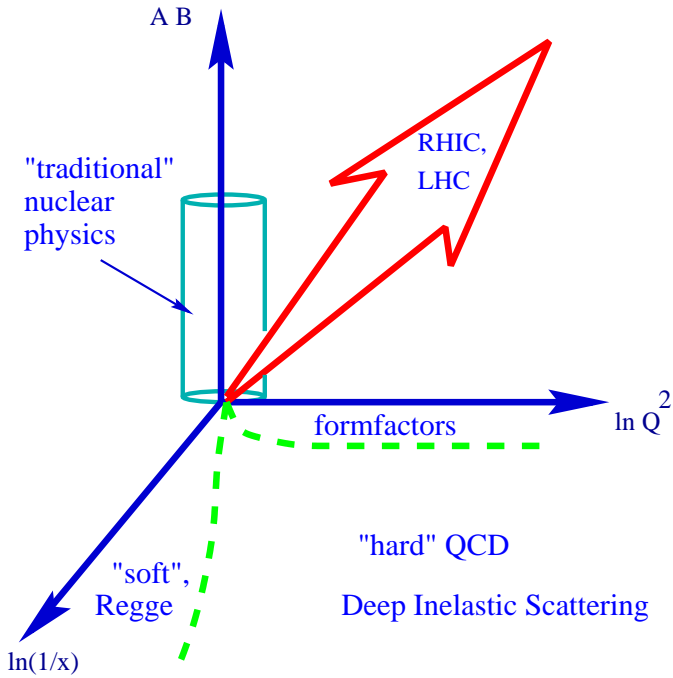}
\caption{The place of relativistic heavy ion physics in the study 
of QCD; the vertical axis is the product of atomic numbers of projectile 
and target, and the horizontal axes are the momentum 
transfer $Q^2$ and rapidity $y = \ln(1/x)$ ($x$ is the Bjorken scaling 
variable).}
\label{fig:rhic}
\end{minipage}
\end{figure}

\section{NEW FRONTIERS OF QCD}

What is the place of relativistic heavy ion program in modern physics? 
 Heavy ion research is aimed at understanding 
QCD, the fundamental particles of which -- quarks and gluons -- are 
already well established; QCD has firmly occupied its place as part of the 
Standard Model. However, 
understanding the physical World does not mean only establishing its 
fundamental constituents; it means, mostly, understanding how these 
constituents interact and bring to the existence the entire variety of 
physical objects 
composing the Universe. Think of electrodynamics -- the simplest of all gauge 
theories -- which is responsible for an enormous assortment of materials and 
substances of different structure. Now try to imagine the beauty and 
complexity 
of collective phenomena made possible in the theory where ``electrons'' carry 
three different ``charges'', ``photons'' carry 
eight, and they are all bound by the force two orders of magnitude stronger 
than electro-magnetic forces! 
Just as the research in condensed matter physics is  
driven by the ability to perform 
experiments with different number of atoms, under different conditions of 
low and high temperature and pressure, further progress in QCD will be largely driven 
by the studies of hadronic matter under extreme conditions. 
By increasing the atomic number of the colliding systems and by raising the energy of 
the collision, we get access to the high parton density, 
high field strength QCD (see Fig.\ref{fig:rhic}).  

\newpage

The heavy ion program thus brings us to the important new frontier 
of modern physics. I believe we are at the beginning of a long and exciting journey.

\end{document}